\documentclass{aa}  

\usepackage{graphicx}
\usepackage{txfonts}

\usepackage{hyperref}
\hypersetup{colorlinks,linkcolor={blue},citecolor={blue},urlcolor={black}}

\DeclareMathAlphabet{\mathbbm}{U}{bbm}{m}{n}

\usepackage{scalerel}

\begin{document} 

\title{The CMB lensing imprint of cosmic voids detected in the WISE-Pan-STARRS luminous red galaxy catalog}

\author{G. Camacho-Ciurana\inst{1}\fnmsep\inst{2}\fnmsep\inst{3} \and P. Lee\inst{1}\fnmsep\inst{2} \and N. Arsenov\inst{4}\fnmsep\inst{1}\fnmsep\inst{2} \and A. Kov\'acs\inst{1}\fnmsep\inst{2} \and I. Szapudi\thanks{MTA Guest Professor 2023}\inst{5}\fnmsep\inst{1} \and I. Csabai\inst{3}}

\institute{MTA-CSFK Lend\"ulet "Momentum" Large-Scale Structure (LSS) Research Group, 1121 Budapest, Konkoly Thege Mikl\'os \'ut 15-17, Hungary
\and Konkoly Observatory, HUN-REN CSFK, MTA Centre of Excellence, Budapest, Konkoly Thege Mikl\'os {\'u}t 15-17. H-1121 Hungary
\and Department of Physics of Complex Systems, ELTE E\"otv\"os Lor\'and University, Pf. 32, H-1518 Budapest, Hungary
\and Institute of Astronomy and NAO, Bulgarian Academy of Sciences, 72 Tsarigradsko Chaussee Blvd., 1784 Sofia, Bulgaria
\and Institute for Astronomy, University of Hawaii, 2680 Woodlawn Drive, Honolulu, HI, 96822, USA
}

\titlerunning{Cosmic voids in WISE-PS1 LRGs $\times$ {\it Planck} CMB lensing}
\authorrunning{G. Camacho-Ciurana et al.}

\date{Received December 2023}

 
\abstract
{The cross-correlation of cosmic voids with the lensing convergence ($\kappa$) map of the Cosmic Microwave Background (CMB) fluctuations offers a powerful tool to refine our understanding of the dark sector in the consensus cosmological model.}
{Our principal aim is to compare the lensing signature of our galaxy data set with simulations based on the concordance model and characterize the results with an $A_{\kappa}$ consistency parameter normalized to unity. In particular, our measurements contribute to the understanding of the "lensing-is-low" tension of the $\Lambda$CDM model.}
{In this analysis, we selected luminous red galaxies (LRGs) from the WISE-Pan-STARSS data set, allowing an extended cross-correlation measurement using 14,200 deg$^2$ sky area, that offers a more precise measurement compared to previous studies. We created 2D and 3D void catalogs to cross-correlate their locations with the {\it Planck} CMB lensing map and studied their average imprint signal using a stacking methodology. Applying the same procedure, we also generated a mock galaxy catalog from the WebSky simulation to serve as a basis for comparison.}
{The 2D void analysis revealed good agreement with the standard cosmological model with $A_{\kappa}\approx1.06 \pm 0.08$ amplitude, i.e. $S/N=13.3$, showing a higher signal-to-noise than previous studies using voids detected in the Dark Energy Survey (DES) data set. The 3D void analysis exhibited a lower signal-to-noise ratio and demonstrated worse agreement with our mock catalog than the 2D voids. These deviations might be attributed to limitations in the mock catalog, such as imperfections in the LRG selection, as well as a potential asymmetry between the North and South patches of the WISE-Pan-STARSS data set in terms of data quality.}
{Overall, we present a significant detection of a CMB lensing signal associated with cosmic voids, largely consistent with the concordance model. Future analyses using even larger data sets also hold great promise of further sharpening these results, given their complementary nature to large-scale structure analyses.}
\keywords{cosmic microwave background, gravitational lensing}
\maketitle
%

\section{Introduction}
\label{sec:Section1}

\emph{Cosmic voids} are large under-dense structures occupying most of the late-time Universe. Unlike other components of the cosmic web (walls, filaments and galaxy clusters), void interiors are less prone to non-linear gravitational effects since they contain significantly less dark matter \citep[see e.g.][for a recent review]{Pisani2019}. As a consequence, voids are dominated by dark energy and therefore these regions are becoming promising laboratories for extracting cosmological information. In particular, they offer novel ways to test the $\Lambda$CDM (Lambda-Cold Dark Matter) model, modified gravity scenarios \citep[see e.g.][]{Clampitt2013,Cai2015,Cautun2018,Baker2018,Schuster2019,Davies2021}, or to constrain the neutrino mass due their sensitivity to diffuse components \citep{Kreisch2019,Contarini2021}.  

Voids constrain cosmological models through various probes, e.g. the void size function, density and velocity profiles, and also their evolution with redshift \citep[see e.g.][]{pisani2015,Verza2019,Nadathur2020,Aubert2020,Hamaus2021}. Moreover, their lensing signals serve as complementary tools for exploring the underlying dark matter distribution, showing a de-magnification effect photons are deflected while traversing cosmic voids. We note, however, that the detection of cosmic shear or convergence signal from an individual void is challenging due to the important uncertainties \citep[][]{Amendola1999,Krause2013}, detections of the void lensing signal using \emph{stacking} methods from large catalogues of voids have already been reported \citep{Melchior2014,Sanchez2016,Gruen2016,ClampittJain2015,Brouwer2018,Fang2019,Jeffrey2021}, including lensing analyses of given different void definitions and simulation with different cosmological models \citep[see e.g.][]{Cautun2016,Davies2018,Davies2021}.

To further probe the properties of dark matter and dark energy, an alternative strategy is to stack the Cosmic Microwave Background (CMB) on the positions of cosmic voids and thus measure their imprints in temperature or lensing convergence maps. Along these lines, various studies have probed the integrated Sachs-Wolfe (ISW) effect \citep{SachsWolfe} by stacking the temperature anisotropy maps on void positions. See also \cite{alonso18} and \cite{Li2024} for explorations of void signals in Compton y-maps to study the thermal Sunyaev-Zeldovich (tSZ) effect.

This tiny CMB foreground signal has generated considerable interest, as the $A_{\scaleto{\rm ISW}{4pt}}$ = $\Delta T_\mathrm{obs}/\Delta T_{\scaleto{\rm \Lambda CDM}{4pt}}$ amplitude parameter (the ratio of observed and expected ISW signals) has often been found significantly higher from $R\gtrsim100~h^{-1}\mathrm{Mpc}$ voids, or \emph{supervoids}, than expected in the concordance $\Lambda$CDM model \citep[see e.g.][for discussions about the expected ISW signal from voids, and on the role of selection effects]{Nadathur2012,Flender2013,Ilic2013,HM2013}. Relying first on Wilkinson Microwave Anisotropy Probe \citep[][WMAP]{WMAP9} and then \emph{Planck} CMB data \citep[][]{Planck2018_cosmo}, moderately significant excess $A_{\scaleto{\rm ISW}{4pt}}$ amplitude values \citep{Granett2008,Cai2017,Kovacs2018} and good consistency with the $\Lambda$CDM model predictions \citep{Hotchkiss2015,NadathurCrittenden2016} have both been reported, using the Sloan Digital Sky Survey (SDSS) luminous red galaxy (LRG) catalogs, and than Baryon Oscillation Spectroscopic Survey (BOSS) data to construct void catalogs. The enhanced ISW signals from supervoids are also considered anomalous because 2-point correlation analyses do \emph{not} show significant excess either, compared to $\Lambda$CDM predictions \citep[see e.g.][]{PlanckISW2015,Stolzner2018,Hang20212pt}.

We note that the ISW excess problem has also been linked to the CMB \emph{Cold Spot} anomaly \citep{CruzEtal2004}, and significant evidence exists for the presence of the low-$z$ \emph{Eridanus} supervoid in its direction \citep[see e.g.][]{SzapudiEtAl2014,Kovacs2022}. It was shown, however, that assuming the $\Lambda$CDM model, the size and under-density of the aligned supervoid is not sufficient to fully explain the observed temperature depression \citep{Nadathur2015,Naidoo2016,Mackenzie2017}.

Further complicating the picture, similar excess ISW signals were also reported using the Dark Energy Survey \citep[][DES]{DES} LRG catalogs \citep{Kovacs2016,Kovacs2019} and the extended Baryon Oscillation Spectroscopic Survey (eBOSS) quasar data set \citep{Kovacs2021}. 

These results have also generated interest in supplementing the ISW analyses with the CMB lensing signal of voids, since the convergence maps provide different but related information about the evolution of the underlying gravitational potentials. Voids correspond to local minima in the lensing convergence ($\kappa$) maps, estimated from the matter density field $\delta(r,\theta)$ via projection as
\begin{equation}
\kappa(\theta)=\frac{3H_0^2\Omega_m}{2c^2}
    \int_{0}^{r_{\rm max}} \delta (r,\theta)
    \frac{(r_{\rm max}-r)r}{r_{\rm max}}\, dr
    \label{eq:kappa_born}
 \end{equation}
in the Born approximation with the Hubble constant $H_{\mathrm{0}}$ and matter density parameter $\Omega_{\mathrm{m}}$, where $r$ denotes the co-moving distance to source galaxies in the background of the lenses (with distorted shapes due to lensing), and $r_{\rm max}$ determines the maximum distance considered.

Using BOSS data, \cite{Cai2017} and \cite{Raghunathan2019} both detected the CMB lensing imprint of voids, finding good consistency with the simulation-based $\Lambda$CDM expectations ($A_{\kappa}=\kappa_{\scaleto{\rm data}{4pt}}/\kappa_{\scaleto{\rm model}{4pt}}\approx1$).
Measurements using the DES LRG catalogs, however, reported a moderately low $A_{\scaleto{\rm \kappa}{4pt}}\approx0.82\pm0.08$ amplitude \citep{Vielzeuf2019,Kovacs2022_DES}, that is also consistent with the findings by \cite{Hang2021} who analysed both voids and superclusters detected in the Dark Energy Spectroscopic Instrument \citep[][DESI]{Dey2019} Legacy Survey galaxy data set, finding an overall $A_{\scaleto{\rm \kappa}{4pt}}\approx0.81\pm0.06$ best-fit amplitude.

In this paper, we perform a novel analysis and identify cosmic voids from the WISE-Pan-STARRS galaxy data set to better understand previously reported tensions concerning the ISW and lensing imprints of these large-scale structures in the \emph{Planck} maps. The paper is organized as follows. In Section \ref{sec:Section2}, we introduce our mock and observational data sets.  Then, Section \ref{sec:Section3} contains a description of our stacking methodology and error analysis. We then present our main observational results in Section \ref{sec:Section4}, followed by a summary of our main conclusions in Section \ref{sec:Section5}.

\section{Data sets}
\label{sec:Section2}
\subsection{WISE-Pan-STARRS galaxies}

As tracers of the cosmological large-scale structure (LSS), we used the \emph{cross-matched} catalogue of galaxies from the all-sky Wide-Field Infrared Survey Explorer 
\citep[WISE,][]{Wright2010} and 3$\pi$ Panoramic Survey Telescope and Rapid Response System (Pan-STARRS1, PS1 for short) sky surveys presented by \citet{Beck2022}. The PS1 Data Release 2 (DR2) catalogs provide broad-band photometric measurements (Kron and PSF magnitudes) of about three quarters of the sky using the $g, r, i, z, y$ filters in the optical range \citep[see][for further details]{tonry, panstarrs, Magnier2020, Magnier2020b, Magnier2020c, Waters2020}. The WISE survey scanned the full sky in four infrared photometric bands (W1, W2, W3, W4) having effective wavelengths of 3.4, 4.6, 12 and 22 $\mu$m, respectively. Regarding the high noise level and the relatively large number of missing error estimates of the W3, W4 filters, we only considered WISE measurements obtained in the W1 and W2 filters. We then only used W1 information to apply selection cuts among the WISE-PS1 galaxies. 

Our analysis applied further selection cuts to the WISE-PS1 galaxy catalog and focused on a subset of LRGs. While their exact 3D position information is not accessible using photo-$z$ data, the expected $\sigma_{z}\approx0.02(1+z_{\rm spec})$ precision allows a robust reconstruction of the largest cosmic voids \citep[see e.g.][for further details]{Sanchez2016}. We defined a sample of WISE-PS1 LRGs using the following color and magnitude cuts, inspired to some extent by the LRG target selection strategy \citep{Zhou2023} applied by the Dark Energy Spectroscopic Instrument \citep[DESI,][]{DESI} survey:
\begin{itemize}
    \item $r-z > (z-17.3)/2$ (\emph{PS1 color cut})
    \item $z<20.7$ (\emph{PS1 magnitude cut})
    \item $i-W1 > 0.5$ (\emph{WISE-PS1 color cut})
    \item $ext\_flg\neq0$ (\emph{WISE extended object cut})
    \item $cc\_flags='0000'$ (\emph{WISE cut for flagged data})
\end{itemize}

\begin{figure*}
\begin{center}
\includegraphics[width=170mm]{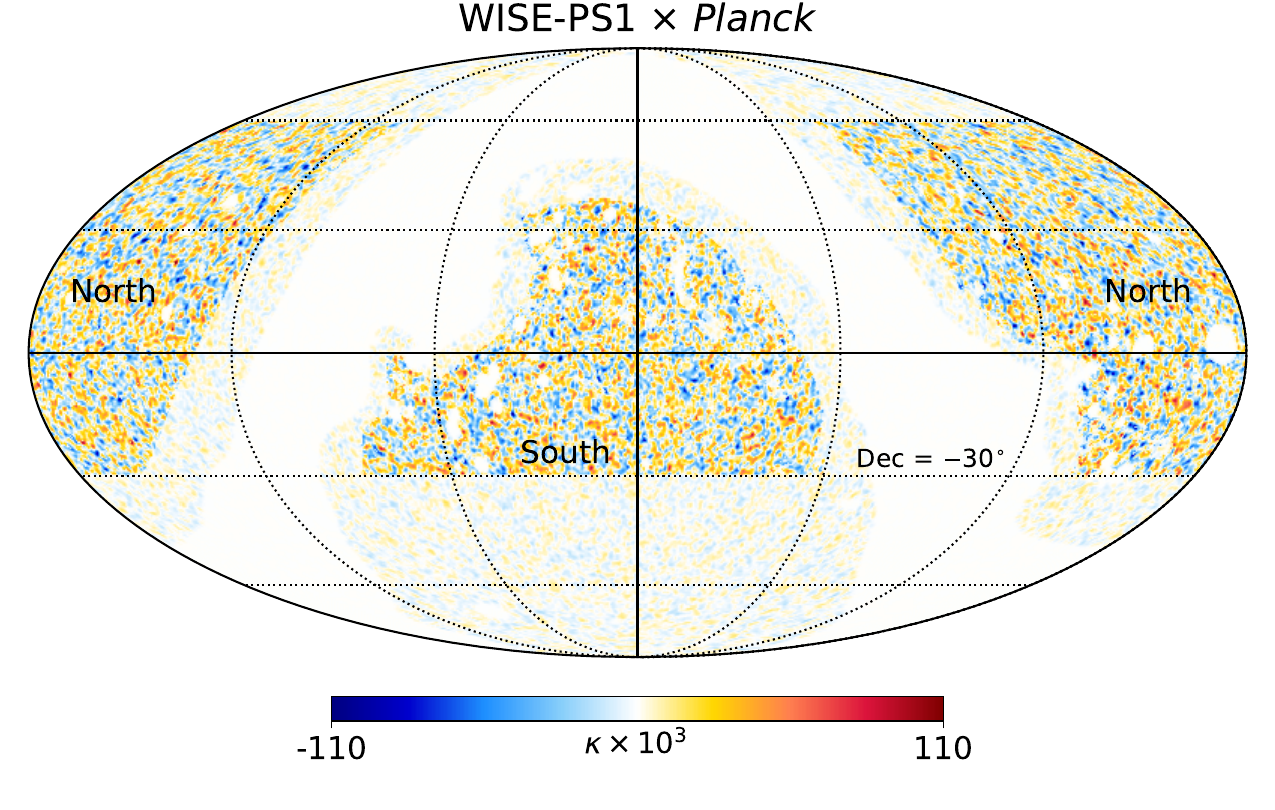}
\caption{\label{fig:figure_1} The {\it Planck} CMB lensing map (with a FWHM=$1^{\circ}$ Gaussian smoothing) is highlighted within the WISE-PS1 sky area, where we performed a robust measurement of voids $\times$ CMB $\kappa$ cross-correlations. The data is divided into two larger patches, labelled as North and South, comprising an unmasked area of approximately $A=14,200$ $\rm deg^{2}$. }
\end{center}
\end{figure*}

Observed magnitudes have been corrected for the galactic dust extinction using the related extinction coefficients ($\alpha_g = 3.172$, $\alpha_r = 2.271$, $\alpha_i = 1.682$, $\alpha_z = 1.322$, $\alpha_y = 1.087$, $\alpha_{W1} = 0.18$) and the $E(B-V$) dust extinction values of a map that is based on PS1 observations of Milky Way stars \citep{Schlafly2014}. 

We also applied an additional $0.42<z_{\rm phot}<0.7$ photometric redshift cut to focus on the most dense part of the WISE-PS1 data with the highest galaxy density, where the resulting sample did not show any visually obvious contamination or inhomogeneity. 

Finally, we created a sky-mask to minimize both Galactic and extra-galactic sources of possible contamination and biases. Our \texttt{healpix} mask \citep{Healpix}, created at $N_{\rm side}=512$ resolution, was constructed by the following process:
\begin{itemize}
    \item we applied the point source mask based on the {\it Planck} 2018 measurements \citep{Planck2018}
    \item dusty pixels with E(B-V) > 0.1 were removed from the analysis (see \cite{SchlegelEtal1998} for further details)
    \item we removed pixels with $|b|<25^{\circ}$, i.e. close to the Milky Way
    \item pixels near the North Galactic cap were removed ($b>60^{\circ}$)
\end{itemize}

The above LRG selection cuts and sky-masking strategy resulted in a sample of about $N_{g}=2.46\times10^6$ WISE-PS1 galaxies, distributed in an unmasked area of approximately $A=14,200$ $\rm deg^{2}$ i.e. covering about $1/3$ of the sky (see Figure \ref{fig:figure_1} for details). 

\subsection{CMB lensing map from the \emph{Planck} mission}

In our LSS$\times$CMB cross-correlation measurements, we relied on the reconstructed CMB lensing convergence ($\kappa$) map provided by the {\it Planck} collaboration. This data set is released in the form of $\kappa_{\rm lm}$ spherical harmonic coefficients \citep[see][for details]{Planck2018_lensing} up to  $\ell_{\rm max}=2048$. We created a $\kappa$ map at $N_{\rm side}=512$ resolution by converting the $\kappa_{\rm lm}$ values into \texttt{healpix} maps. We also used a corresponding mask from the publicly available {\it Planck} data (part of our analysis mask described above). In Figure \ref{fig:figure_1}, we show the common WISE-PS1 survey area on top of the {\it Planck} $\kappa$ map, indicating our conservative choices to avoid possibly contaminated regions. Additionally, the Dec > $-30^{\circ}$ PS1 data-taking limit is also marked.

We note that, while higher resolution maps can be constructed from the available $\kappa_{\rm \ell m}$ coefficients, the chosen $N_{\rm side}=512$ is a sufficient choice given the degree-size angular scales of cosmic voids in our analysis. We also applied a FWHM=$1^{\circ}$ Gaussian smoothing to the convergence map to suppress small-scale noise patterns, following the S/N optimisation results by \cite{Vielzeuf2019}. Another filtering we implemented was removing the large-scale modes from the $\kappa$ map ($\ell<20$) to remove any remnant wide-angle systematic effect in the map. Our analyses consistently applied these smoothing and filtering steps to the observed and simulated $\kappa$ maps.

Finally, we also removed a slight remnant bias from the $\kappa$ map ($\bar{\kappa}\approx -10^{-4}$), which is the result of using the WISE-PS1 sky mask instead of the \emph{Planck} mask such that the construction provides a zero-mean convergence map.

\subsection{Simulations: mock galaxies and mock CMB $\kappa$ map}

To create a simulated data set for calibrating our pipeline, we analysed the publicly available WebSky simulation\footnote{\url{https://mocks.cita.utoronto.ca/data/websky/}} \citep{websky2}, that was constructed assuming a {\it Planck} 2018 cosmology \citep{Planck2018_cosmo}. Based on the peak-patch method \citep[i.e. not a full N-body run, see][for further detail]{websky1}, the WebSky simulation encloses a vast total cosmic volume of $V\simeq 1,900$~$({\rm Gpc})^3$ with mass resolution of about $M_{\rm{min}} \simeq 1.2 \times 10^{12} M_\odot$ at $0<z<4$. It provides a full-sky light-cone catalogue that contains about $9\times 10^{8}$ dark matter halos (each with RA, Dec, redshift, an initial Lagrangian position [Mpc], final Eulerian position [Mpc], velocity [km/s], and mass [$M_{200\overline{\rho}_m}$]).

Notably, another key objective of the WebSky simulation was to provide a set of secondary CMB foreground maps correlated with the simulated halo catalogs. Therefore, to model the WISE-PS1 $\times$ {\it Planck} cross-correlations, we also used the WebSky CMB lensing convergence map in our analyses, that was constructed using the field particles and a Navarro-Frenk-White (NFW) profile \citep[NFW,][]{NFW1996} for all halos in the peak-patch halo catalogue that subtend more than a pixel. \cite{websky2} added another component to the convergence, as an uncorrelated Gaussian random field, to account for the power originating from redshifts larger than the maximum included in the simulation, z$_{\rm{max}}$=4.5.

Naturally, this signal-only WebSky CMB lensing map is insufficient to account for observational noise in the {\it Planck} $\kappa$ reconstructions. In the following section, we provide details on how we added {\it Planck}-like noise to estimate the measurement errors, which ultimately limit the precision of the void lensing profiles.

\section{Methodology}
\label{sec:Section3}

Our main objective is to measure the imprint of cosmic voids on CMB lensing convergence maps, and below we detail our main tools. We closely follow the methodology by \cite{Vielzeuf2019} and \cite{Kovacs2022_DES}, building on their findings regarding e.g. optimisation techniques and free parameters in the analysis.

\subsection{Void finder methods}

To complement standard LSS-CMB cross-correlation methods using clusters or filaments \citep[see e.g.][]{Baxter2015,Madhavacheril2015,He2017,Baxter2018}, the under-dense regions of the Universe might also be identified by void finder algorithms. Identifying cosmic voids is a complex process, affected by survey properties including tracer quality, tracer density, and masking effects \citep[see e.g.][]{Sutter_bias, Nadathur2015}. In particular, void properties depend significantly on the methodology used to define the voids and thus working with different void finders make the analysis more robust. In this study, we employed two distinct void finding methods that are standard methods in the field \citep[see e.g.][for details]{Fang2019}. 

\subsubsection{Cosmic voids in 2D maps}

Our main analysis pipeline uses a 2D void finder method, developed by \cite{Sanchez2016} for the analysis photo-$z$ data sets. The algorithm works by projecting galaxies in two-dimensional redshift slices specified by the user. We followed the original methodology \citep[see also][for more recent applications]{Vielzeuf2019} and chose a thickness of $100~h^{-1}\mathrm{Mpc}$, which corresponds to about twice the size of typical photo-$z$ errors. Then, we projected the data into \texttt{healpix} maps with $\text{N}_{\text{\rm side}} = 512$ resolution in 7 redshift slices in the $0.42<z<0.7$ range covered by the WISE-PS1 LRGs.

The next step is a Gaussian smoothing applied to the projected galaxy count maps, and voids are defined by \emph{minima} of the smoothed tracer density field in the given slice \citep{Sanchez2016}. Annuli of increasing radii are defined around the prospective void centers, and the algorithm stops when mean density within the slice ($\delta$ = 0) is reached in a given annulus, where the void radius ($R_v$) is defined. Void catalogs produced by the above algorithm depend on two main free parameters that we set to the following values:

\begin{itemize}
    \item Smoothing scale ($\sigma$): this parameter defines the Gaussian smoothing scale applied to the density map to define density minima. Here we employed a $\sigma=10~h^{-1}\mathrm{Mpc}$, also used by e.g. \cite{Kovacs2022}, which enables a reliable identification of larger voids that carry most of the detectable lensing signal (without smoothing out the density troughs too much, which would reduce the void lensing signal's amplitude in the centre).
    \item  Under-density threshold ($\delta_{\rm min}$): in a smoothed density map, this parameter defines the minimum threshold to consider a given local density minimum as a possible void centre. Naturally, significant voids correspond to lower $\delta$ values, and we again followed \cite{Vielzeuf2019} by setting $\delta_{\rm min}=-0.2$ to eliminate shallow under-densities.
\end{itemize} 

\subsubsection{Cosmic voids in 3D}

To compare with 2D voids detected from the same galaxy catalogs, we used the \texttt{REVOLVER} (REal-space VOid Locations from surVEy Reconstruction) void-finding code\footnote{\url{https://github.com/seshnadathur/Revolver}} \cite{Nadathur2019} to also construct a 3D voids catalog. The algorithm is based on the \texttt{ZOBOV} technique \citep{ZOBOV} to reconstruct the local density of tracers, employing a Voronoi tessellation and subsequently identifying density minima by comparing Voronoi cells with their neighboring cells. 

Presumably, the analysis of 3D voids is less efficient in detecting lensing signals than using 2D voids, which guarantee longer photon paths in a coherent gravitational potential \citep[see e.g.][]{Fang2019,Davies2021}. Yet these alternative void definitions and sub-classes may also offer important checks of the fiducial 2D voids analyses.

Moreover, voids detected using the \texttt{ZOBOV} method also provide a useful proxy for the sign of the gravitational potential in and around voids, defined as
\begin{equation}
\label{eq:lambda_v}
\lambda_v\equiv\overline\delta_g\left(\frac{R_{\mathrm{eff}}}{1\;h^{-1}\mathrm{Mpc}}\right)^{1.2}
\end{equation}
using the average galaxy density contrast, $\overline\delta_g = \frac{1}{V}\int_{V}\delta_g\,\mathrm{d}^3\mathbf{x}$, and the effective spherical radius, $R_\mathrm{eff}= \left(\frac{3}{4\pi}V\right)^{1/3}$, where the volume $V$ is determined from the sum of the volumes of Voronoi cells making up the void \citep[see e.g.][]{NadathurCrittenden2016}.

We note that two consecutive runs of \texttt{ZOBOV} pipeline do not result in identical void catalogs. While differences are small in total the number of voids and in the main void parameters, in our analysis we created 10 slightly different versions of our WISE-PS1 and WebSky 3D void catalogs, and used their median signals in our CMB lensing cross-correlation measurements for more robust results.

\begin{figure}
\begin{center}
\includegraphics[width=82mm]{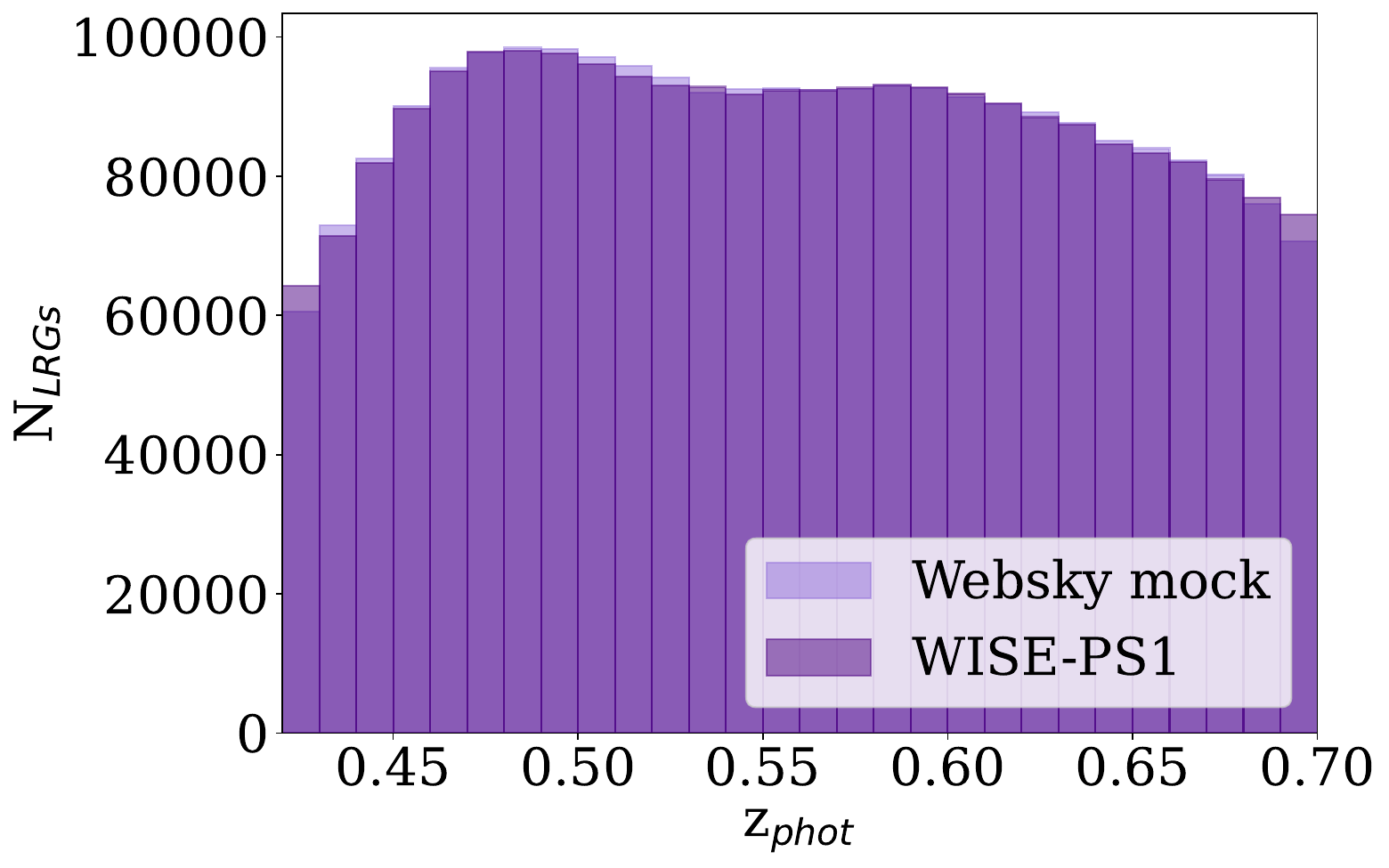}
\caption{\label{fig:figure_2} Comparison of the redshift distribution of LRGs in the WISE-PS1 data set versus in the WebSky mock. We followed an HOD modelling approach to create a realistic model of the source density as a function of redshift, reaching a few-per-cent match throughout the whole range.}
\end{center}
\end{figure}

\begin{figure*}
\begin{center}
\includegraphics[width=165mm]{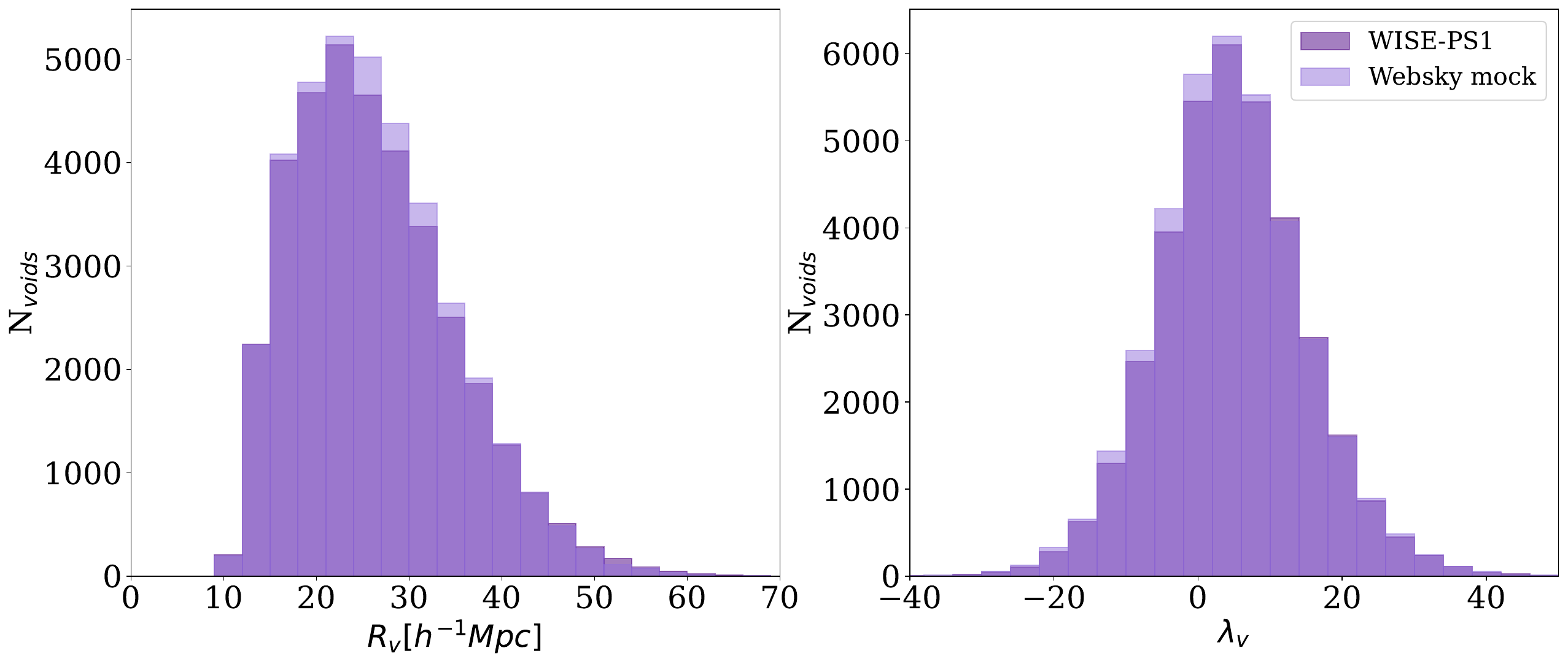}
\caption{\label{fig:figura_3D_voidspar}
Void properties in the WISE-PS1 data set and in the WebSky mock catalog. The void radii (left) and $\lambda_v$ parameter (right) both show great agreement, which, after further pruning steps, facilitates a detailed comparison of the simulated and observed lensing signals from these voids.}
\end{center}
\end{figure*}

\subsection{Modelling void properties with mock LRGs}

Considering the realistic modelling of the WISE-PS1 LRG sample of galaxies, we followed a halo occupation distribution (HOD) methodology \citep[see e.g.][]{Tinker2012}, using the publicly available \texttt{pyHOD} code\footnote{\url{https://github.com/Yucheng-Zhang/pyhod}}. The HOD approach describes the average number of central and satellite galaxies residing in halos as a function of halo mass, $M$. The expected total number of galaxies in a dark matter halo is the sum of the central and satellite galaxy probabilities, expressed as
\begin{equation}
\langle N_\mathrm{tot}(M) \rangle = \langle N_\mathrm{cen}(M) \rangle + \langle N_\mathrm{sat}(M) \rangle.
\end{equation}

The probability that a dark matter halo, typically a massive one ($M_{\rm{min}} \simeq 10^{13} M_\odot$ halo mass), contains a central LRG is given by the smooth step function
\begin{equation}
\langle N_\mathrm{cen}(M) \rangle = \frac{1}{2} \left[1 + \mathrm{erf} \left( \frac{\log M - \log M_\mathrm{min}}{\sigma_\mathrm{\log M}} \right) \right],
\end{equation}
where the position of this step is set by $M_\mathrm{min}$. A halo with mass $M \ll M_\mathrm{min}$ hosts no central galaxy,
which then transitions according to the value of the $\sigma_\mathrm{\log M}$ parameter to $P=1$ probability for $M \gg M_\mathrm{min}$ halos \cite[see e.g.][for a recent application]{Zhou2020}. 

The number of satellite galaxies in each halo is Poisson distributed, with a mean value given by a power law,
\begin{equation}
\langle N_\mathrm{sat}(M) \rangle = \left( \frac{M-M_\mathrm{0}}{M_\mathrm{1}} \right)^{\alpha},
\label{eq:hod_satellite_power_law}
\end{equation}
with $\alpha$, $M_\mathrm{0}$, and $M_\mathrm{1}$ as free parameters of the satellite occupation model, where satellite galaxies are randomly positioned in the halo following an NFW profile.

We then tuned the free parameters of the HOD model. Our main objective was to make the void parameters in the simulated WebSky catalogs entirely consistent with those of the WISE-PS1 voids.

\begin{figure*}
\begin{center}
\includegraphics[width=165mm]{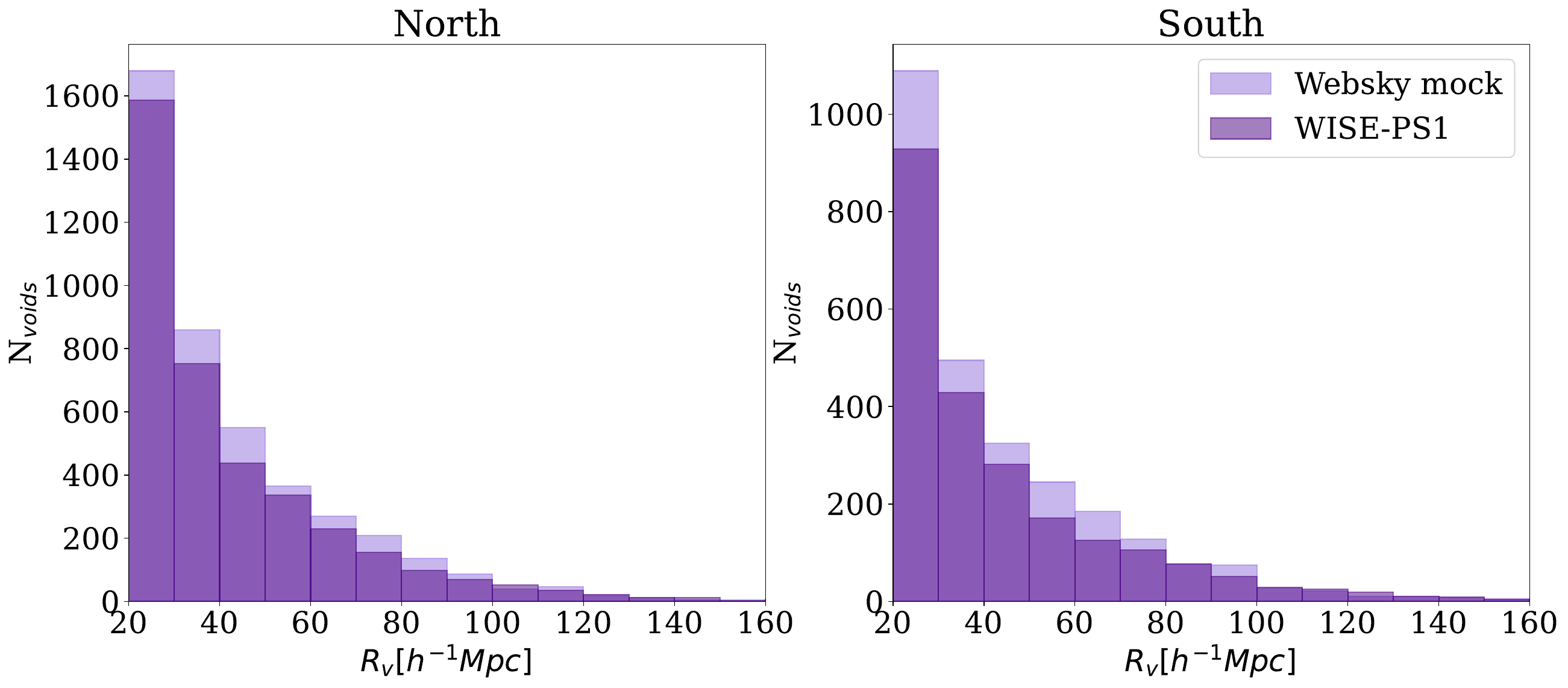}
\caption{\label{fig:figure_3} The number of voids present in the WISE-PS1 and WebSky mock catalogs is compared in the North (left) and South (right) regions in the figure. The difference in the coverage areas of the North and South masks accounts for the variation in the number of voids observed. Furthermore, there is a 13\% difference in the overall number of voids between the LRG catalogs, with 6090 voids in the WISE-PS1 catalog and 6992 voids in the WebSky mock catalog. We use the filter definition $R_v >20 h^{-1}\text{Mpc}$, $\delta_{c}$<-0.3 and $\delta<-0.05$, following \cite{Sanchez2016}.}
\end{center}
\end{figure*}

We followed the DESI LRG target selection strategy \citep{Zhou2023} in our overall strategy to populate the most massive halos with mock LRGs. In particular, we found that splitting the halo catalogs to three redshift bins and setting the following HOD parameters results in the best match between the WISE-PS1 data and the WebSky simulation:
\begin{itemize}
    \item $0.4<z<0.55$: $\log M_\mathrm{min}=12.87$, $\log M_\mathrm{0}=13.67$
    \item $0.55<z<0.65$:  $\log M_\mathrm{min}=12.98$, $\log M_\mathrm{0}=13.78$
    \item $0.65<z<0.76$:  $\log M_\mathrm{min}=13.06$, $\log M_\mathrm{0}=13.69$
\end{itemize}

Further HOD parameter values were identical in the three redshift bins (with marginal influence on the results), with $\log M_\mathrm{1}=13.9$, $\sigma_\mathrm{\mathrm{\log M}}=0.3$, and $\alpha=1.29$. 

As a final step, we added Gaussian photo-$z$ errors to the accurately known simulated galaxy redshifts, using a $\sigma_{z}\approx0.02(1+z_{\rm spec})$ scatter that is realistic for the WISE-PS1 LRGs \citep{Beck2022}. When applying the WISE-PS1 sky mask on the full-sky WebSky simulation, we populated the halo distribution with about $N_{g}=2.46\times10^6$ mock galaxies, that is in close agreement with the observed WISE-PS1 data set, as demonstrated in Figure \ref{fig:figure_2}. 

Our best-fit version of mock catalog of 3D voids is shown in Figure \ref{fig:figura_3D_voidspar}. We note that our choice to choose 3D voids as a basis of mock-data comparison was motivated by the availability of more details than in 2D catalogs. The satellite fraction of the resulting WISE-PS1-like mock LRG sample is about $5\%$ with $f_\mathrm{sat}\approx0.05$. We found that alternative versions of the mock LRG catalog with higher satellite fractions ($10\%, 20\%$) resulted in slight mismatches in the void radius and under-density distribution, and this baseline $5\%$ choice guaranteed the best agreement between mock and data.

We then also compared the properties of 2D voids in the WebSky and WISE-PS1 data sets. In Figure \ref{fig:figure_3}, we show a comparison of the observed and simulated 2D void catalogs, where the average radius for both catalogs is about $R_v\approx 40 h^{-1}\text{Mpc}$, with similar minimum and maximum values. 

We carried out the void finding process separately in the North and South patches, in order to minimize any possible large-scale contamination effect from different survey data quality in the two main survey areas. Yet, the agreement in 2D voids is less perfect compared to the 3D voids, and we first detected a $20\%$ mismatch between WISE-PS1 and WebSky in the total number of 2D voids. We then performed a few tests to increase the level of the agreement (prior to looking at the lensing signals), including the removal of the large-scale modes ($\ell<20$) from the galaxy density maps. We found that their absence reduces the disparity in the number of 2D voids to approximately $13\%$, suggesting that remnant large-scale systematic effects might still be present in the WISE-PS1 data set. We note that the typical size of a void is $\sim 1^\circ$, thus the lensing measurements are not significantly affected by this filtering.

\subsection{Stacking measurements}

Studying cosmic voids individually can be challenging as their lensing signals are weak and therefore noisy in real-world measurements \citep{Krause2013}. To address this issue, an established technique is to use stacking methods \citep[see e.g.][]{Granett2008,Kovacs2015, Hang2021}, which nonetheless also requires a large number of voids to reduce the observational noise and detect a clear void lensing signal.

\begin{figure*}
\begin{center}
\includegraphics[width=170mm]{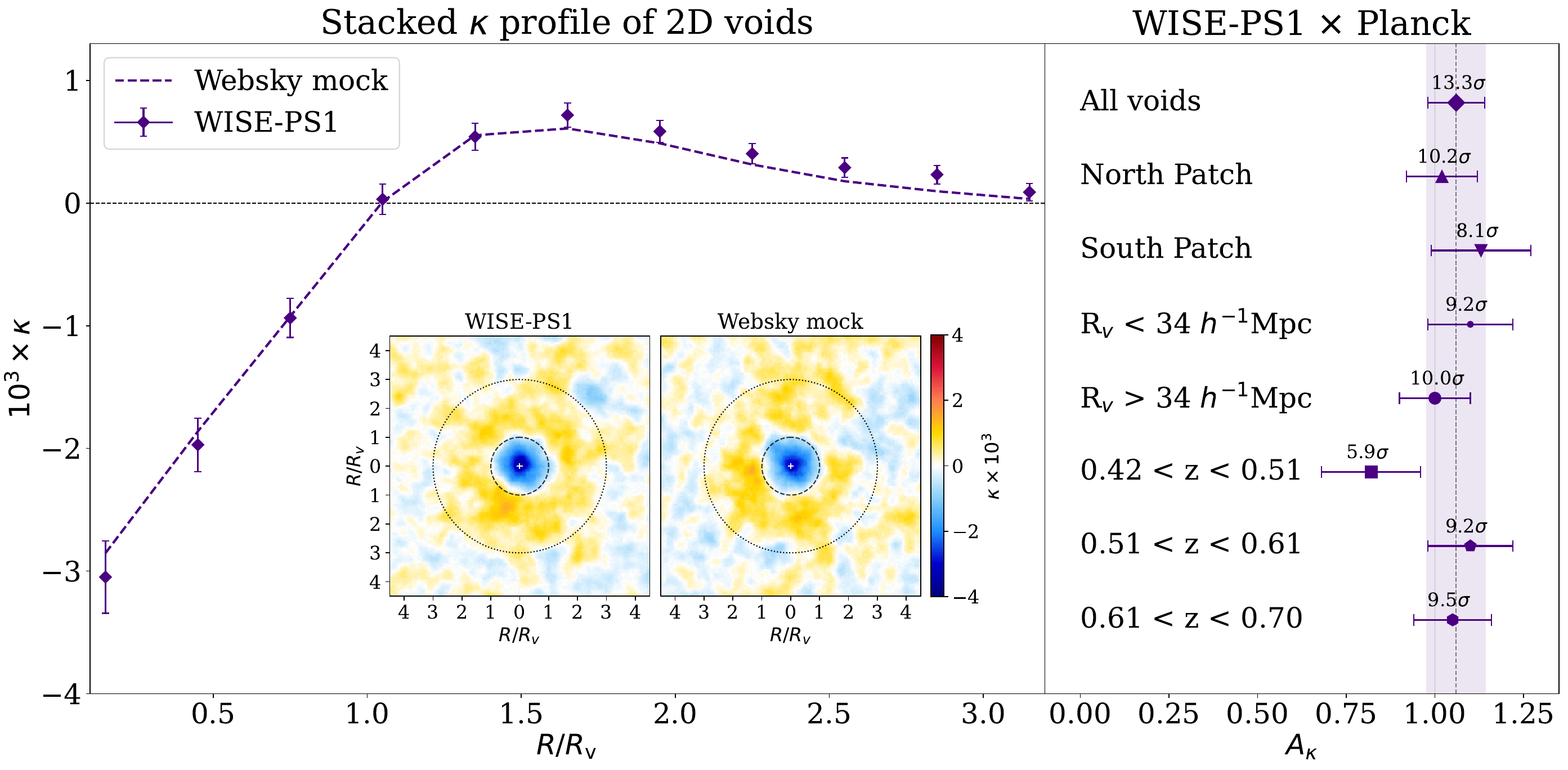}
\caption{\label{fig:figura_2D_voids} \textit{Left.} Stacked CMB $\kappa$ radial profile and image of the Websky mock and WISE-PS1 catalogs using all 2D voids. The $\kappa$ profiles (with 11 radial bins) show great agreement, with an 6\% higher than expected signal at the center, and a clear negative $\kappa$ value inside of the voids, followed by a clear detection of the void edges in the lensing signal too. We also find good agreement when looking at the stacked images, where a random noise map realisation was added to the WebSky mock for a more realistic impression. \textit{Right.} Here we show the best-fit $A_\kappa$ parameter values for the fiducial all-voids case (A$_\kappa=1.05\pm 0.08$, i.e. S/N=13.1), and we also present our separate the North and South results. We also show how splitting the void catalogs into two halves at R$=34h^{-1}$Mpc also shows consistent signals. Finally, we also present our main findings when splitting the voids into three redshift bins using $\triangle z\approx 0.1$ width, with a weak trend for stronger signals at higher redshifts.}
\end{center}
\end{figure*}

The stacking method uses square-shaped cut-outs from the CMB $\kappa$ map aligned with the center of each void, followed by a relative re-scaling of cut-out images by the known angular size of the voids. Any subsequent statistic is then expressed in the re-scaled $R/R_v$ units (see the inset in Figure \ref{fig:figura_2D_voids} for example). 

Beyond the stacking image, one may also measure the radial profile with the number of bins and the corresponding resolution as free parameters. We set $\text{N}_{\text{bin}}=11$ and up to $(R/R_v)=3.5$ (i.e. significantly beyond the void radius to also capture environmental effects), resulting in $\Delta (R/R_v)=0.3$ bin width in our analysis.

\subsection{Error analysis}
\label{subsec:error_analysis}

The most significant source of uncertainty in the CMB lensing signal of the stacked cosmic voids comes from the instrumental noise in the \emph{Planck} data. In addition, the total error has subdominant contributions from overlap effects of multiple voids along the same line-of-sight, adding an intrinsic variation to the measurable signal. 

To factor both error sources, we created 2000 \emph{Planck}-like noise realizations ($N_{\kappa}^i$) and 2000 WebSky-like CMB $\kappa$ map realizations ($S_{\kappa}^i$). Based on the study by \citealt{Vielzeuf2019}, we randomly produced the noise maps using the power spectrum released by \citealt{Planck2018_cosmo} with the \texttt{synfast} routine. Then, we extracted the power spectrum of the WebSky CMB lensing map  using the \texttt{anafast} routine, and generated random signal maps that are \emph{uncorrelated} with the voids. These maps are then added to the \emph{Planck}-like noise maps ($N_{\kappa}^i$+$S_{\kappa}^i$). Finally, all the 2000 random maps are stacked on the positions of the voids to characterize the covariance of our cross-correlation measurements.

During this procedure we note that the maps were pre-processed before the stacking measurements, using the same methodology as in the case of the \emph{Planck} $\kappa$ map. We removed the fluctuations at the largest scales in the map ($\ell<20$) and applied a Gaussian smoothing with $FWHM=1^\circ$. 

Following the above steps, the level of consistency between WISE-PS1 and WebSky becomes quantifiable, which results in constraints for the best-fitting CMB lensing amplitude parameter $A_{\kappa}=\kappa_{\scaleto{\rm WISE-PS1}{4pt}}/\kappa_{\scaleto{\rm WebSky}{4pt}}$ and its corresponding uncertainty ($\sigma_{A_{\kappa}}$). To measure this parameter, we followed the analysis steps by \citealt{Vielzeuf2019}, evaluating the statistic
\begin{equation}
    \chi^2 = \sum_{ij} (\kappa_i^{\rm WISE-PS1}-A_{\kappa}\cdot\kappa_i^{\rm WebSky})C_{ij}^{-1}(\kappa_j^{\rm WISE-PS1}-A_{\kappa}\cdot\kappa_j^{\rm WebSky})
\end{equation}
where $\kappa_i$ is the average CMB lensing signal in a bin radius $i$ and $C$ is the covariance matrix. By minimizing the $\chi^2$ values, we are able to constrain the best-fitting $A_{\kappa}\pm \sigma_{A_{\kappa}}$, which characterizes the agreement between the WISE-PS1 and WebSky signals.

\section{Results \& Discussion}
\label{sec:Section4}

In this section, we compare the lensing signal from the WISE-PS1 and WebSky void catalogs, using 2D and 3D methods. In both analyses, we worked separately in the North and the South patches (see Figure \ref{fig:figure_1}), using a simple weighting approach to generate the $\kappa$ signal of the entire sky area, based on the number of voids in each patch. With this procedure, we aimed to determine the level of agreement between the WISE-PS1 data and the \emph{Planck}-like $\Lambda$CDM cosmology used in the WebSky simulation (where an $A_{\kappa}\approx 1$ indicates a good concordance).

\subsection{2D voids}

Following the above methodology for the 2D void finding algorithm, we detected 6090 voids in the WISE-PS1 data, and we found 6992 voids in the WebSky mock within the same sky area ($14,200$ $\rm deg^2$).

In Figure \ref{fig:figura_2D_voids}, we present the main results for the 2D void catalogs. We show the stacked images and the reconstructed lensing profiles for all voids detected in the WISE-PS1 data set and in the WebSky mock catalog, as well as the best-fit lensing amplitude results for different bins of redshift, radii, and sky patches. 

\begin{figure*}
\begin{center}
\includegraphics[width=170mm]{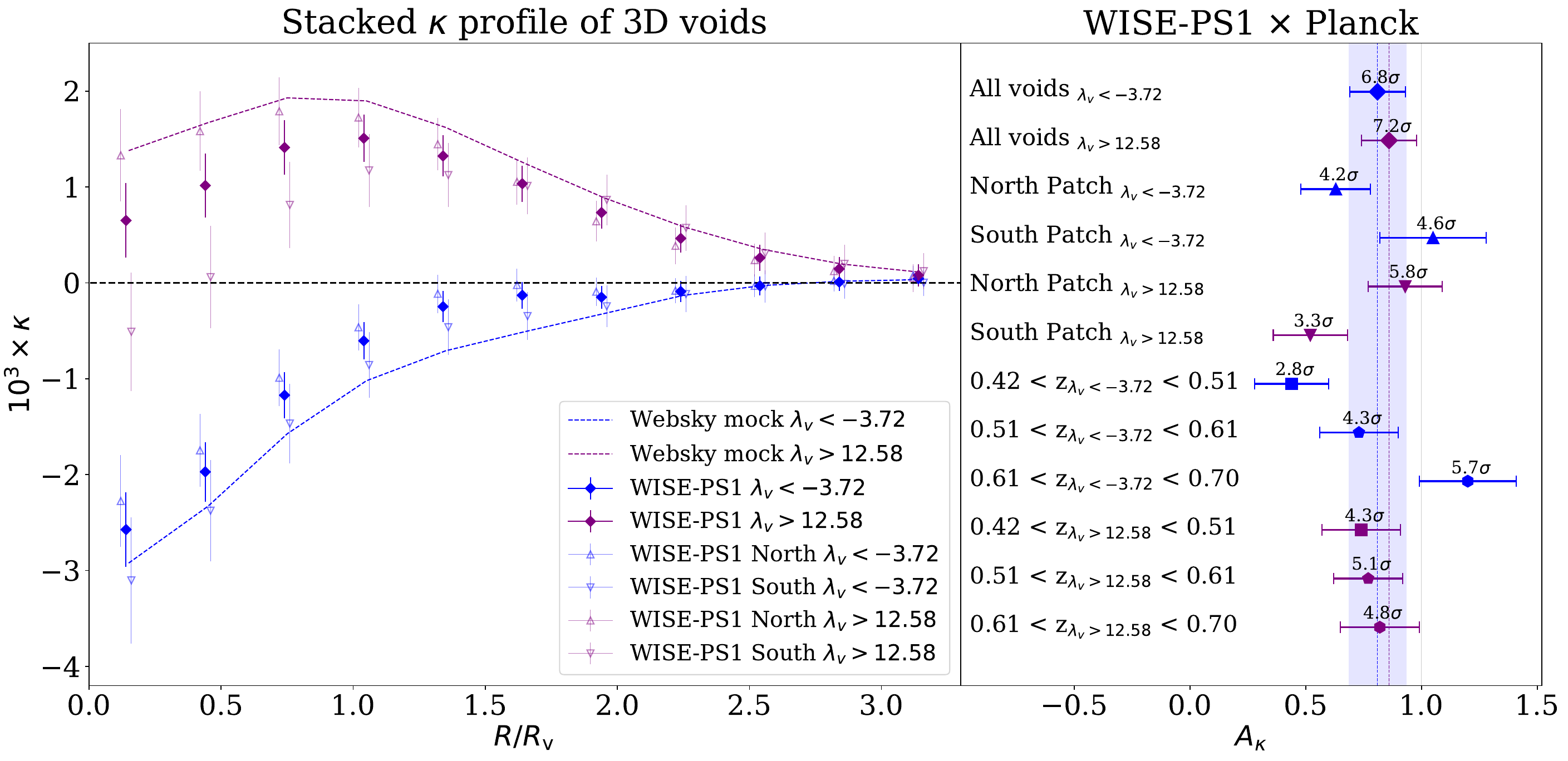}
\caption{\label{fig:figura_3D_voids} \textit{Left.} Stacked CMB $\kappa$ profiles measured for 3D voids in the WebSky and the WISE-PS1 data sets. We generally split the all-voids sample into voids-in-voids ($\lambda_v<-3.72$) and voids-in-clouds ($\lambda>12.58$) bins, which are expected to carry most of the observable void lensing signal among all $\lambda_v$ values. \textit{Right.} We show the best-fit $A_\kappa$ parameters for different cases. The main results are $A_\kappa\approx0.81\pm 0.12$ for voids-in-voids and A$_\kappa\approx0.86\pm 0.12$ for voids-in-clouds, both with $S/N\approx 7$.}
\end{center}
\end{figure*}

As expected, we found a clear negative lensing signal in the void interiors, although the signal of WISE-PS1 voids is about $8\%$ higher in the centre than the WebSky-based expectations. At $\rm R/R_v>1$ the $\kappa$ signal starts to be positive (where the compensating over-dense regions is located), marking the edge of the voids. Far from the centre, the $\kappa$ signal of both the observational and simulated profiles converge to zero. This trend is also visible in the stacked images, where we added a random noise map realization to the WebSky mock image, in order to look more similar to the observational WISE-PS1 image.

To calculate the agreement between WebSky and WISE-PS1 voids catalogs, we estimated the best-fit value of the $A_{\kappa}$ parameter, that describes the ratio of the observed and the expected void lensing profiles. For our fiducial result using all voids, we obtained $A_{\kappa}\approx1.06 \pm 0.08$, i.e. a $\textup{S/N} \approx 13.3$ detection of a cross-correlation signal between voids and CMB lensing convergence.

For further insights, we then performed a similar analysis for different subsets of the void catalogs. As illustrated in Figure \ref{fig:figura_2D_voids}, the $A_{\kappa}$ values measured in the North and South WISE-PS1 patches are consistent with each other. We also applied a radius binning by splitting the void catalog into two identical halves at $R=34 h^{-1}$Mpc. The results again show good consistency between the larger and smaller voids. Finally, we also split the $0.42<z<0.7$ void catalog into three redshift bins, with $\triangle z \approx 0.1$. In this test, we found a weak trend with stronger void lensing signals at the higher redshift bins, but overall all results are broadly consistent with $A_{\kappa}\approx1$, as demonstrated in Figure \ref{fig:figura_2D_voids}. For numerical values of the best-fit lensing amplitudes, see Table \ref{tabla:comparison} below.

\subsection{3D voids}

Considering the 3D void catalog, we make further selection cuts on the $\lambda_v$ parameter to study the specific sub-classes of voids. Typically surrounded by other voids, the voids-in-voids exhibit $\lambda_v<0$ values, while the voids-in-clouds tend to be embedded in generally over-dense environments with $\lambda_v>0$, i.e. these two classes also correspond to distinct CMB lensing imprints \citep[see][for further details]{Raghunathan2019}. We note that shallow and small voids are expected to produce weaker lensing signals than the ones with more extreme $\lambda_v$ values, and the associated noise is also greater due to their smaller angular size. 

Therefore, we chose $\lambda_v<-3.72$ to study the most significant 20\% of voids-in-voids, or under-compensated void class, while we set $\lambda_v>12.58$ to also define a voids-in-clouds (over-compensated) sample, with 7430 voids in both subsets. The remaining 60\% of the 3D voids with lower absolute $\lambda_v$ values are excluded from our analysis.

The summary of these results is depicted in Figure \ref{fig:figura_3D_voids}. Unlike the good consistency with the $A_\kappa \approx1$ expectations for 2D voids, we found that from the 3D void analysis, most $A_\kappa$ values are below unity. For voids-in-voids ($\lambda_v<-3.72$ subset), we obtained $A_{\kappa}\approx0.81\pm0.12$, i.e. S/N=6.8, that is broadly consistent with our best-fit amplitude in the voids-in-clouds regime ($\lambda_v>12.58$) with $A_{\kappa}\approx0.86\pm0.12$ (S/N=7.2). We also noticed, however, that the North and the South patches for these two void types show substantial differences. The South patch for the $\lambda_v>12.58$ sub-sample shows $A_{\kappa}\approx0.52\pm0.15$ (i.e. a rather low S/N=3.3), indicating possible remnant systematic effects in the input data sets. 

As in the case of the 2D voids, we again implemented a redshift-binning approach using the same bin edges and width. For the voids-in-voids sub-class, we observed a trend with $A_{\kappa}$ values increasing with redshift, with higher significance than for the 2D results. However, we do not observe this apparent trend for voids-in-clouds ($\lambda_v>12.58$), which suggests that this feature is again a consequence of remnant systematic effects in the data.

\begin{table}
\caption{Results from the error analysis for the 2D and the 3D voids with different binning, comparing the A$_\kappa$, signal-to-noise and number of voids.}
\label{tabla:comparison}
\centering                          
\begin{tabular}{c c c c}
\hline\hline                 

2D voids & $A_{\kappa} \pm \sigma_{\textup{A}_{\kappa}}$ & S/N & $\textup{N}_{\textup{voids}}$ \\
\hline
All  & $1.06\pm0.08$ & 13.3 & 6090 \\
North Patch & $1.02\pm0.10$  & 10.2 & 3812 \\
South Patch & $1.13\pm0.14$  & 8.1 & 2278 \\
$R_v<34h^{-1}\textup{Mpc}$ & $1.10\pm0.12$  & 9.2 & 2936 \\
$R_v>34h^{-1}\textup{Mpc}$ & $1.00\pm0.10$  & 10.0 & 3154 \\
0.42 < z < 0.51 & $0.82\pm0.14$  & 5.9 & 1998 \\
0.51 < z < 0.61 & $1.10\pm0.12$  & 9.2 & 1800 \\
0.61 < z < 0.70 & $1.05\pm0.11$  & 9.5 & 2292 \\ 
\hline\hline
3D voids & $A_{\kappa} \pm \sigma_{\textup{A}_{\kappa}}$ & S/N & 
$\textup{N}_{\textup{voids}}$ \\
\hline
All low $_{\lambda_v<-3.72}$ & $0.81\pm0.12$ & 6.8 & 7430 \\
All high $_{\lambda_v>12.58}$ & $0.86\pm0.12$ & 7.2 & 7430 \\
         
North Patch $_{\lambda_v<-3.72}$ & $0.63\pm0.15$  & 4.2 & 4749 \\
South Patch $_{\lambda_v<-3.72}$ & $1.05\pm0.23$  & 4.6 & 2681 \\
         
North Patch $_{\lambda_v>12.58}$ & $0.93\pm0.16$  & 5.8 & 4572 \\
South Patch $_{\lambda_v>12.58}$ & $0.52\pm0.16$  & 3.3 & 2858 \\

0.42 < z$_{\lambda_v<-3.72}$ < 0.51 & $0.44\pm0.16$  & 2.8 & 2279 \\
0.51 < z$_{\lambda_v<-3.72}$ < 0.61 & $0.73\pm0.17$  & 4.3 & 2724 \\
0.61 < z$_{\lambda_v<-3.72}$ < 0.70 & $1.20\pm0.21$  & 5.7 & 2427  \\

0.42 < z$_{\lambda_v>12.58}$ < 0.51  & $0.74\pm0.17$  & 4.3 & 2512 \\
0.51 < z$_{\lambda_v>12.58}$ < 0.61  & $0.77\pm0.15$  & 5.1 & 2537 \\
0.61 < z$_{\lambda_v>12.58}$ < 0.70  & $0.82\pm0.17$  & 4.8 & 2381 \\
\hline\hline                                  
\end{tabular}
\end{table}

\section{Conclusions}
\label{sec:Section5}

In this paper, we extended the series of CMB lensing $\times$ voids cross-correlation measurements. Our motivation was to learn more about the dark matter content of cosmic voids by reducing the statistical uncertainties, and to possibly learn about the validity of the puzzling "lensing-is-low" anomaly \citep[see e.g.][]{Heymans2021}. 

We analyzed a WISE-PS1 LRGs sample with 14,200 deg$^2$ sky area, allowing a more precise cross-correlation measurement than achieved in most previous studies. We cross-correlated both, 2D and 3D void catalogs with the {\it Planck} lensing map, and thus studied their average $\kappa$ imprint using a stacking methodology, relying on a mock galaxy catalog as reference, built from the WebSky dark matter simulation \citep[][]{websky1}.

Our fiducial analysis revealed good agreement with the standard $\Lambda$CDM, finding an $A_{\kappa}\approx1.06 \pm 0.08$ lensing amplitude for 2D voids, i.e. $S/N\approx13.3$ (see Figure \ref{fig:figura_2D_voids}). We did not find significant evidence for deviations from this main results in the North vs. South patches of the WISE-PS1 catalog, splitting into large vs small voids, or when binning the voids into 3 redshift slices (see Table \ref{tabla:comparison}). We report that our detection is of higher $S/N$ than previous studies using voids detected in the DES Year-3 data set \citep[see e.g.][]{Vielzeuf2019,Kovacs2022_DES}, and it is comparable to DESI Legacy Survey results \citep[][]{Hang2021}. 

The 3D void analysis exhibited lower $S/N$ and it agrees less precisely with the WebSky mock than our findings based on 2D voids. We found $A_{\kappa}\approx0.81\pm0.12$ amplitude for the voids-in-voids subset ($\lambda_v<-3.72$) and $A_{\kappa}\approx0.86\pm0.12$ for voids-in-clouds ($\lambda_v>12.58$), i.e. about $1\sigma$ lower than the expected $A_{\kappa}\approx1$ (see Figure \ref{fig:figura_3D_voids}). However, we detected even lower amplitudes when we analyzed the North and South patches separately, and especially when we analyzed different redshift bins in the case of the voids-in-voids sample (see Table \ref{tabla:comparison}). These findings lead us to the conclusion that, despite the lack of similar trends in the fiducial 2D void analysis, our WISE-PS1 LRG catalog might feature remnant systematic effects that affect the outcomes.

The observed mild deviations from $A_{\kappa}\approx1$ might be attributed to limitations in our simulation approach, or to imperfections in the WISE-PS1 LRG selection. We thus conclude that the explanation of the observed moderately significant $2-3\sigma$ tensions in some of our binned void lensing signals may well be observational effects, rather than cosmological in their nature.

Overall, we presented a significant detection of a CMB lensing signal associated with about 7000 cosmic voids at $0.42<z<0.7$, largely consistent with the concordance model. Future analyses using even larger data sets (e.g. \emph{Euclid}, Vera C. Rubin Observatory, Roman, SPHEREx, J-PAS, etc) will provide further insights into cosmological tensions, and to the nature of voids in particular.

\section*{Data availability}

The WISE-PS1 galaxy catalogues and masks are publicly available as part of the PS1 DR2 release (using freely available WISE data for the cross-matching), as well as the WebSky simulation, and also the \emph{Planck} 2018 CMB lensing map. Void catalogs and CMB cross-correlation measurement code will be made available upon reasonable request to the corresponding author.

\section*{Acknowledgments}

The Large-Scale Structure (LSS) research group at Konkoly Observatory has been supported by a \emph{Lend\"ulet} excellence grant by the Hungarian Academy of Sciences (MTA). This project has received funding from the European Union’s Horizon Europe research and innovation programme under the Marie Skłodowska-Curie grant agreement number 101130774. Funding for this project was also available in part through the Hungarian Ministry of Innovation and Technology NRDI Office grant OTKA NN129148. The authors thank Conor McPartland and R\'obert Beck for their help with the general WISE-PS1 galaxy catalogs that we used to select LRGs for our analysis.

\bibliographystyle{aa}
\bibliography{refs}

\end{document}